\documentclass[12pt]{article}
\usepackage{color}
\usepackage{amsfonts,amssymb,amsmath}
\usepackage[export]{adjustbox}
\usepackage{graphicx}
\usepackage[T1]{fontenc}
\usepackage[numbers,sort&compress]{natbib}
\graphicspath{ {./images/} } \textheight 9in \textwidth  6.5in
\begin{document}
\title{\textsf{Adiabatic charging of open quantum battery}}
\author{M. A. Fasihi\thanks{Email: a.fasihi@gmail.com},
\hspace{2mm}R. Jafarzadeh Bahrbeig\thanks{Email:
r.jafarzadeh86@gmail.com},\hspace{2mm}B. Mojaveri
\thanks{Email: bmojaveri@azaruniv.ac.ir; bmojaveri@gmail.com}, and R. Haji Mohammadzadeh\thanks{Email: www.ranamohammadzadeh7@gmail.com}\\
{\small {Department of Physics, Azarbaijan Shahid Madani University,
PO Box 51745-406, Tabriz, Iran \,}}} \maketitle
\begin{abstract}
We revisit the adiabatic charging of a three-level QBs, using the
adiabatic quantum master equation formalism. We restrict ourselves
to the weak-coupling regime with an Ohmic thermal bath and
investigate the effects of relaxation and dephasing on the charging
process. We analyze the dependence of the stored energy, ergotropy
as well as efficiency of QB on the total time of evolution $t_f$. We
demonstrate that for very short charging time ($t_f$), where the
evolution is highly non-adiabatic, the stored energy and ergotropy
are very small. However, with increasing $t_f$ we show that there is
an optimal charging time, $t_f^{opt}$, for maximum energy charging
such that at low temperatures we could fully charge the battery and
effectively extract the whole amount of energy from it. Note that,
the optimal charging time could be decreased by adjusting strength
of the coupling between system and environment and also appropriate
choice of the Hamiltonian parameters which in turn speed up the
charging process. On the other hand, we show that for very long
charing time $t_f$ the charging energy, ergotropy and efficiency
decrease due to thermal excitations. Furthermore to get more
insights about the problem we investigate the distance between
density matrix of system at optimal charging time $t_f^{opt}$ and
the corresponding thermal state using one-norm distance.
\end{abstract}
\section{introduction}
The emergence of quantum batteries, as a new topic of quantum physics and quantum technology, is one of the results of recent interest and research to the quantum thermodynamics \cite{Vinjanampathy-2016,Alicki-2018,Deffner-2019}. Alicki and Fannes in \cite{Alicki-2013} put forward the notion of QB as a generalization of its classical counterpart. In fact, QBs are quantum devices to store and extract the energy to consume it in other quantum devices, as do their counterpart in the classical world. Indeed, QB usually can be modeled by a single qubit \cite{Andolina-2018,Giov-2019,Santos-2020}, or ensemble of qubits \cite{Le-2018,Wang-2019,Rossini-2019,Kamin-2020,Crescente-2020,Munro-2020}. It is worth to remark that, the qubit-based QBs are charged through different protocils such as charger-mediated process where QB interacts directly with a charger which could be a quantum system such as a qubit system or a quantum harmonic oscillator \cite{Andolina-2018,Giov-2019}. Environment-mediated charging is the other charging protocol where QB interacts with the charger mediated with and environment \cite{Tabesh-2020,Song-2022,Morrone-2023,Xu-2024}. Recently charging process of QBs based on non-equilibrium steady state of spin system have been investigated \cite{Mojaveri-2024}. Also some charging protocols exploit external fields such as optical fields \cite{Ghosh-2021} and parity-deformed fields \cite{Mojaveri1-2024}. Moreover, to improve the performance of QBs some interesting protocols based on adjusting the velocity of battery and charger qubit have been proposed \cite{Mojaveri-2023}. Although the recent researches on QBs provide very important improvements, there are some serious obstacles toward the realization of QBs. Among them unstable charging and discharging is the most important one.
Indeed, for stabilization of charged state, different quantum control strategies based on feedback control such as linear feedback control \cite{Mitchison-2021}, homodyne-based feedback control \cite{Yao-2022,Koshihara-2023} have been proposed. Furthermore, the techniques based on a sequence of repeated quantum measurements \cite{Pasquale-2017,Gherardini-2019} have been used in \cite{Gherardini-2020} to stabilize open QB. However, according to the method of repeated measurement one need to access the battery continuously which leads to the energy consumption.
We also remark that, charging protocols in most of the qubit batteries are unstable, because in the qubit-based batteries the energy will go back and forth between charged and uncharged states \cite{Andolina-2018,Le-2018,Wang-2019}. Thus the charging as well as discharging performance highly depends on the precise control over the interaction time. In fact, to avoid discharging the battery after its full charging, it is necessary to decouple battery from charger or external field, which in practice is a complicated task as it explained in \cite{Santos-2019}. However, Santos and his collaborators proposed a qutrit system as an alternative candidate to build the QB in order to achieve efficient charging and discharging protocols. Indeed, interaction of a three-level atom with quantized light, give rise to some interesting and important phenomena in quantum optics such as inversionless lasing \cite{Scully-1989,Mompart-2000} and stimulated Raman adiabatic passage \cite{Gaubatz-1990,Vitanov-2017,Shore-2017}. The latter feature is essential for efficient deriving of qutrit quantum system, which is not possible in the case of two-level ones.

On the other hand, in practice, quantum devices including QBs generally must be considered as open quantum systems \cite{Giov-2019,Yao-2022,Liu-2019}, because the real quantum systems interact with their surrounding environment. Recently, in \cite{Santos-2019}, an interesting scheme introduced  based on the adiabatic time evolution and Raman adiabatic stimulated passage to stabilize energy storage in an open and closed single three-level QB. Indeed, based on this technique, by exploiting two external fields, it is possible to transfer energy coherently between the ground state and the most upper excited state along dark state, thereby coherently charging a three-level battery. More interestingly, this idea \cite{Santos-2019} is realized experimentally with superconducting circuits \cite{Santos-2022}. Moreover, the results have been confirmed with the most recent Study \cite{R.Zheng-2022}, where Zheng et al proposed a method to control dynamics of three-level open systems and realized it in the experiment with a superconducting qutrit. In addition, in \cite{Y.Zheng-2023}, the frequency-modulated stimulated Raman adiabatic passage technique theoretically and experimentally have been used to improve the charging (discharging) efficiency of a cascaded three-level QB that is constituted of a superconducting transmon qutrit. This viewpoint has also been confirmed in \cite{Munro-2020}, where the authors, applied dark state to stabilize the stored energy of the multiple two-level QB. Meanwhile, motivated by the idea of \cite{Santos-2019} and its realization \cite{Santos-2022} several similar protocols have been proposed for efficient stable and fast charging, such as \cite{Gemme-2024,Yang-2024,Dou-2020,Dou-2022}.

In this paper, inspired by \cite{Santos-2019,Santos-2022} and following the idea of adiabatic master equation \cite{Albash-2012} we motivated to revisit the problem of adiabatic charging of open three-level QB and analyze it in the frame of adiabatic master equation. It is worth to remark that, according to adiabatic master equation \cite{Albash-2012}, for an open quantum system with a time dependent Hamiltonian, the corresponding Lindblad operators are time dependent which is quite different from \cite{Santos-2019} and the followed literatures. Exploiting adiabatic master equation in weak coupling regime and using dark state, we investigate the effects of relaxation, dephasing and different parameters of system and environment on the charging process and its stabilization in a three-level quantum battery. We study the dependence of the final stored energy and final ergotropy of battery on total time of system evolution $t_f$. However, it is very important to note that, the role of $t_f$ in adiabatic charging of open QB is quite different in comparison with that of closed-battery. Indeed, in the closed-system setting, the only relevant time scale is the condition that the evolution be sufficiently adiabatic, i.e., $T_{ad}\sim 1/\Delta_{min}$ (the heuristic adiabatic condition $t_f \gg T_{ad}$ suppresses nonadiabatic transitions and assures that final state reached has high overlap with the ground state of the final Hamiltonian \cite{Kato-1950,Lidar3-2009,Amin1-2009}.) While in the open-system setting, if phase decoherence appears just in the energy eigenbasis, the rest of relevant time scales which determine the efficiency of the charging process are $t_f$, the relaxation time $T_1$, and the time scale $T_{ad}$. Therefore, the efficiency of the adiabatic charging process of an open QB, depends on the interplay between these time scales which is not considered in the recent researches. Indeed, the interplay between these time scales is quite complicated in comparison with that of the closed-system setting. In fact, as we will discuss, the role of $t_f$ is quite ambiguous when the dynamics of open QB is governed via Morkovian adiabatic master equation in the presence of thermal bath. Therefore, despite to closed-system adiabatic dynamic, in adiabatic charging process of an open QB we expect to deal with an optimal value of $t_f$ which is problem dependent \cite{Albash-2012,Steffen-2003,Lidar1-2005,Lidar2-2005,Lidar-2015}. We show that for very short $t_f$, the final stored energy and ergotropy are almost zero. Also we discuss about the effect of environment temperature on the adiabatic charging process of QB. Indeed, at low temperatures, increasing $t_f$ we find an optimal charging time $t_f^{opt}$ for full charging the battery and effective extracting the whole amount of energy from it. We show that the appropriate choice of Hamiltonian parameters leads to the speed up in the charging process. On the other hand, we show that for very long $t_f$s the charging energy, ergotropy and efficiency of the battery decrease. Moreover, using one-norm distance, we determine the distance between density matrix of system and the corresponding thermal state at the same temperature during the time.
\section{Adiabatic master equation}
Adiabatic quantum dynamics since the pioneering work of Born and Fock \cite{Born-1928}, have attract lots of attention. Indeed, because of the  recent theoretical and experimental developments in adiabatic quantum computation \cite{Farhi-2001,Johnson-2011} we have witnessed considerable renewed interests to this old topics of quantum physics in both closed and open quantum systems \cite{Child-2001,Teufel-2003,Sarandy-2005,Alicki-2006,Joye-2007,Huang-2008,Amin-2009,Oreshkov-2010,Salmilehto-2010,Albash-2012}.
In this section we briefly introduce one of the main leading approaches to study the adiabatic evolution of time dependent open quantum systems which put forward by Albash et al. in \cite{Albash-2012}. Based on this approach, in the weak-coupling limit, an adiabatic master equation in Lindblad form for the systems evolution can be derived.  The idea is outlined briefly as follows:

Let us consider the general form of system-bath Hamiltonian as
\begin{eqnarray}
\label{eigenvalues}
  \begin{cases}
H(t)=H_S(t) + H_B + H_I ,\\
H_I= \sum_{\alpha}{g_\alpha A_\alpha \otimes B_\alpha},
  \end{cases}
\end{eqnarray}
where $A_\alpha$ and $B_\alpha$ are respectively the dimensionless Hermitian operators of system and bath, and $g_\alpha$ is the system-bath coupling strength. The time-dependent system Hamiltonian satisfies the following eigenvalue equation
\begin{equation}
H_S(t)|\varepsilon_a(t)\rangle=\varepsilon_a(t)|\varepsilon_a(t)\rangle,
\end{equation}
where, $\varepsilon_a(t)$ and $|\varepsilon_a(t)\rangle$ respectively are the instantaneous eigenvalues and eigenvectors of $H_S(t)$.
The gap between instantaneous ground state, $|\varepsilon_0(t)\rangle$, and exited states, $|\varepsilon_a(t)\rangle (a \geq 1)$, is defined as
\begin{equation}
{\Delta_{min}}\equiv \min_{a,t}(\varepsilon_a(t)-\varepsilon_0(t)) > 0,
\end{equation}
It is worth to remark that, the constrain $\Delta_{min}>0$ determines only those excited states which are different from ground state during the time evolution. Now let us consider the bath correlation functions (we set $\hbar = 1$ from now):
\begin{equation}
\mathcal{B}_{\alpha\beta}(t)=e^{i H_B t} B_\alpha e^{-i H_B t} B_\beta,
\end{equation}
then the characteristic decay time is defined via
\begin{equation}\label{bath correlation time}
|\langle\mathcal{B}_{\alpha\beta}(t)\rangle|=|Tr\big(\rho_B \mathcal{B}_{\alpha\beta}(t)\big)|\sim e^{-t/\tau_B},
\end{equation}
note that $\rho_B$ is the initial state of the bath.

According to \cite{Albash-2012}, for the system evolution, an adiabatic master equation in Lindblad form \cite{Lindblad-1976} can be derived  in the weak coupling limit, where $H_S$ dominates $H_I$, together with standard Born-Markov approximations, rotating wave approximation, and an adiabatic approximation \cite{Albash-2012}:
\begin{equation}\label{weak coupling}
g^2\tau_B \ll \Delta_{min},~~ (\textmd{weak coupling}),
\end{equation}
\begin{equation}\label{Markov condition}
g\tau_B \ll 1,~~ (\textmd{Markov approximation}),
\end{equation}
\begin{equation}\label{Adiabatic condition}
\frac{h}{t_f} \ll \textmd{min}\left\{{\Delta_{min}}^2, {\tau_B}^{-2}\right\},
\end{equation}
where $h=max_{s\epsilon[0,1];~ a,b}|\langle\varepsilon_a(s)|\partial_s H_S(s)|\varepsilon_b(s)\rangle|$ estimates
the rate of change of the Hamiltonian. Inequality (\ref{Adiabatic condition}) results from the combination of heuristic adiabatic approximation with the condition that the instantaneous energy eigenbasis should be slowly varying on the timescale of the bath \cite{Albash-2012}.  Then the master equation in the adiabatic limit with rotating wave approximation (Lindblad form) and weak couling limit is, (for more details please see \cite{Albash-2012}):
\begin{equation}\label{master}
\frac{d}{dt}{\rho(t)}=-i[H_S(t)+H_{LS}(t),\rho(t)]+\mathcal{L}_{wcl}[\rho(t)]
\end{equation}
\begin{equation}\label{LWCL}
\mathcal{L}_{wcl}[\rho(t)]=\sum_{\alpha, \beta}\sum_{\omega}{\gamma_{\alpha \beta}(\omega)\Big(L_{\beta,\omega}(t)\rho(t){L^\dag}_{\alpha,\omega}(t)-
\left\{\frac{1}{2}{L^\dag}_{\alpha,\omega}(t)L_{\beta,\omega}(t),\rho(t)\right\}}\Big),
\end{equation}
where the time-dependent Lindblad operators are
\begin{equation}\label{lindblad operators}
L_{\alpha,\omega}(t)=\sum_{\omega=\varepsilon_b(t)-\varepsilon_{a}(t)}{\langle\varepsilon_a(t)|A_\alpha
|\varepsilon_b(t)\rangle~|\varepsilon_a(t)\rangle\langle\varepsilon_b(t)|}.
\end{equation}
and the sum over $\omega$ is over the Bohr frequencies of $H_s(t)$.
The decay rates
\begin{equation}\label{decayrate}
\gamma_{\alpha\beta}(\omega)=g^2\int_{-\infty}^{\infty}{dt e^{i \omega t}\langle\mathcal{B}_{\alpha\beta}(t)\rangle}
\end{equation}
are Fourier transforms of the bath correlation function and satisfy the KMS
condition \cite{Albash-2012}
\begin{equation}\label{KMS}
\gamma_{\alpha\beta}(-\omega)=e^{-\beta\omega}\gamma_{\beta\alpha}(\omega)
\end{equation}
where $\beta$ is the inverse temperature. Note that, $\gamma_{\alpha\beta}(\omega)$ forms
a positive matrix $\gamma(\omega)$.
Meanwhile, the Lamb shift term is
\begin{equation}\label{HLS}
H_{LS}=\sum_{\alpha\beta}\sum_{\omega}{S_{\alpha\beta}(\omega){L^\dag}_{\alpha,\omega}L_{\beta,\omega}},
\end{equation}
where
\begin{equation}\label{HLS}
S_{\alpha\beta}(\omega)=\int_{\infty}^{\infty}{d\omega'\gamma_{\alpha\beta}(\omega')\mathcal{P}\frac{1}{\omega-\omega'}},
\end{equation}
and $\mathcal{P}$ refers the Cauchy principal value.

\section{Model: time dependent three-level open QB }

Let us consider a bare three-level QB with Hamiltonian $H_0=\sum_{i=1}^{3}{\lambda_i |\lambda_i\rangle\langle \lambda_i|}$, where $|\lambda_3\rangle$ and $|\lambda_1\rangle$ refer to the $\emph{full}$ and the $\emph{empty}$ charged states respectively.
The auxiliary fields can be applied to drive the system and provide transitions between the
energy levels. To this end, we introduce the drive Hamiltonian $H_d(t)$, which in general depends on the structure of the system \cite{{Santos-2019},{Vitanov-2017},{Marangos-1998},{Bergmann-1998},{Demirplak-2003},{Fleischhauer-2005}}, as
\begin{equation}\label{Hs}
H_d(t)= A(t)e^{-i \omega_{12} t}|\lambda_1\rangle\langle \lambda_2|+B(t)e^{-i\omega_{23} t}|\lambda_2\rangle\langle \lambda_3|+ H.c,
\end{equation}
then the system Hamiltonian is $H_S(t) = H_0 + H_d(t)$.
We remark that, in order to analyze the stability and performance of open QB in the presence of the most general effects of environment, we will take into account both relaxation and dephasing phenomena. In fact, these are the most relevent sources of nonunitary dynamics in superconducting circuits \cite{Martinis-2003,Paraoanu-2011,Sillanpr-2012}, which could provide possible platform to realize the introduced QB.
However, for the sake of simplicity it would be better to cancel out $H_0$ by moving the system-bath dynamics to the rotating frame generated by the Hamiltonian $H_0$. Therefore, with this considerations and assuming that the auxiliary fields are in resonance with the energy levels of the battery($H_0$), the Hamiltonian in the rotating frame becomes
\begin{equation}
H(t)=(A(t)|\lambda_1\rangle\langle \lambda_2|+B(t)|\lambda_2\rangle\langle \lambda_3|+ h.c)+ H_B+H_I,
\end{equation}
where $H_I=\sum_{k}{g^x_k} X \otimes(b_k+{b_k}^{\dag})+\sum_{k}{g^z_k} Z \otimes(b_k+{b_k}^{\dag})$. Note that the operators $X$ and $Z$ respectively are the x and z components of spin $1$ operators and $b_k$ and ${b_k}^{\dag}$ are respectively, raising and lowering operators for the $k^{th}$ oscillator with natural frequency $\omega_k$, and $g^x_k(g^z_k)$ are the the corresponding coupling strength to spins $x(z)$.

Therefore, the instantaneous energy eigenvalues and corresponding eigenvectors of the system Hamiltonian $(A(t)|\lambda_1\rangle\langle \lambda_2|+B(t)|\lambda_2\rangle\langle \lambda_3|+ h.c)$ are
\begin{eqnarray}\label{eigenHs}
\begin{cases}
\begin{array}{ccc}
|\varepsilon_1(s)\rangle=\frac{1}{\sqrt{2}}\big(\frac{A(s)}{\Delta(s)}|\lambda_1\rangle-|\lambda_2\rangle+\frac{B(s)}{\Delta(s)}|\lambda_3\rangle  \big),~~~~\varepsilon_1(s)=- \Delta(s)\\
\hspace{0mm}|\varepsilon_2(s)\rangle=\frac{-B(s)}{\Delta(s)}|\lambda_1\rangle+\frac{A(s)}{\Delta(s)}|\lambda_3\rangle,~~~~~~~~~~~\varepsilon_2(s)=0~~~(dark~  state)\\
|\varepsilon_3(s)\rangle=\frac{1}{\sqrt{2}}\big(\frac{A(s)}{\Delta(s)}|\lambda_1\rangle+|\lambda_2\rangle+\frac{B(s)}{\Delta(s)}|\lambda_3\rangle  \big),~~~~\varepsilon_3(s)= \Delta(s)\\
\Delta(s)=\sqrt{A(s)^2+B(s)^2},~~s=t/t_f.\\
 \end{array}
 \end{cases}
\end{eqnarray}
where $t \in [0, t_f ]$.

Considering the eigenvalues $\{\varepsilon_{1}(s)=-\Delta(s),~\varepsilon_2(s)=0,~ \varepsilon_{3}(s)=\Delta(s)\}$ the possible bohr frequency are $\omega=\{0,\pm\Delta(s),\pm2\Delta(s)\}$ and according to eq.(\ref{lindblad operators}) the set of corresponding time dependent Lindblad operators of adiabatic master equation are presented in Eq.(\ref{lindblad operators1}) in appendix. It is worth to remark that, in \cite{Santos-2019}, Santos et al investigate adiabatic charging of three-level open QB using the Lindblad master equation with time independent Lindblad operators. Indeed, in their model the time dependence appears just in the unitary term of lindblad master equation. However, in our protocol based on adiabatic master equation \cite{Albash-2012}, the Lindblad operatos are time dependent which provide new description and insight about adiabatic charging of open QBs. In the following lines we will present some basic definitions about charging process of QBs.
\subsection{Average energy and Ergotropy}

Here we briefly review the definition of average stored energy and ergotropy which are the two main features of a QB. The average transformed energy into a QB at time $t$ is determined as
\begin{equation}\label{averag}
\Delta E=Tr\{\rho_B(t) H_B\}-Tr\{\rho_B(0) H_B\}.
\end{equation}
On the other hand, the term ergotropy which coined by Allahverdyan et al. in \cite{Allahverdyan-2004},
refers to the maximum amount of work that can be extracted from a quantum system (QB) by means of a cyclic unitary process and defined as
\begin{equation}\label{ergo}
W(t)=Tr\{\rho_B(t) H_B\}-min_{U_{B}}Tr\{U_B\rho_B(t)U^\dagger_{B} H_B\},
\end{equation}
note that, the minimization in the definition of ergotropy is taken over all the unitary transformations $U_{B}$ acting
locally on the $\rho_B(t)$.
If we order the eigenvalues of $H_B =\sum_{i=1}^{3}{\lambda_i |\lambda_i\rangle\langle \lambda_i|}$ in increasing
order, $\lambda_1 < \lambda_2 < ... < \lambda_N$, and the eigenvalues of
$\rho_B(t)=\sum_{i=1}^{3}{r_i |r_i\rangle\langle r_i|}$ in decreasing order, $r_1 \geq r_2 \geq ... \geq r_N$,
then the ergotropy of $\rho_B(t)$ is given \cite{Allahverdyan-2004} by
\begin{equation}
W(t)=\sum_{jk}r_j \lambda_k\big(|r_j\lambda_k|^2-\delta_{jk}).
\end{equation}
Meanwhile, we define the efficiency of QB, denoted as $\eta(t)$, which determines how effectively we can exploit the stored energy of QB.  The efficiency of a QB is defined as
\begin{equation}
\eta(t)=\frac{W(t)}{\Delta E}.
\end{equation}
\section{Adiabatic charging of open QB}
Exploiting the adiabatic quantum master equation we investigate the effects of dissipation, dephasing and also impact of different parameters of quantum system(QB) and environment on the charging process of an open three-level QB. Furthermore, we study the dependence of the stored energy, and ergotropy as well as the efficiency of QB on total time of system evolution $t_f$. However, it is very important to note that the role of the
total charging time $t_f$ in adiabatic charging of open-battery is quite complicated because of the existence of different relevant time scales in open QB. Therefore as it explained in the introduction, there is an optimal value of $t_f$ which is correspond to the maximum stored energy and maximum ergotropy of QB, we remark that the optimal value of $t_f$ is problem dependent \cite{Albash-2012,Steffen-2003,Lidar1-2005,Lidar2-2005,Lidar-2015}.

Now let us consider the instantaneous eingenvectors $|\varepsilon_i(s)\rangle$ and $|\varepsilon_j(s)\rangle$, to determine the adiabatic limit for this case, we will use the heuristic adiabatic condition eq.(\ref{Adiabatic condition}):
\begin{equation}\label{ad1}
max \frac{|\langle\varepsilon_j(s)|\partial_s H|\varepsilon_i(s)\rangle|}{t_f \Delta(s)^2}\ll 1,~~
0\leq s\leq 1
\end{equation}
For the rest of calculation, we assume a linear interpolation in $H_S(t)$ i.e., we set $A(s)=\omega_A (1-s)$ and $B(s)=\omega_B s$. For the case of $(i=1,j=2)$ the numerator of eq.(\ref{ad1}) would be equal to $|\langle\varepsilon_2(s)|\partial_s H|\varepsilon_1(s)\rangle|=\frac{\omega_A \omega_B}{\sqrt{2}\Delta(s)}$ and its maximum is $max|\langle\varepsilon_2(s)|\partial_s H|\varepsilon_1(s)\rangle|=\frac{\omega_A \omega_B}{\sqrt{2}\Delta_{min}}$. Clearly the minimum of the instantaneous energy gap $\Delta(s)$ is $\Delta_{min}=\frac{\omega_A \omega_B}{\sqrt{\omega_A^2+\omega_B^2}}$, which is reached at $s = s_{min}=\frac{\omega_A^2}{\omega_A^2+\omega_B^2}$. On the other hand, the min of the denominator of eq.(\ref{ad1}) is $t_f \Delta_{min}^2$.
Putting these altogether we find that
\begin{equation}\label{ad1-1}
max \frac{|\langle\varepsilon_2(s)|\partial_s H|\varepsilon_1(s)\rangle|}{t_f \Delta(s)^2}=\frac{(\omega_A^2+\omega_B^2)^{\frac{3}{2}}}{t_f\sqrt{2}~\omega_A^2\omega_B^2}\ll 1,~~
(0\leq s\leq 1).
\end{equation}
Therefore, this yields for the adiabatic condition
\begin{equation}\label{ad1-2}
t_f \gg \frac{(\omega_A^2+\omega_B^2)^{\frac{3}{2}}}{\sqrt{2}~\omega_A^2\omega_B^2}.
\end{equation}
Note that there is a symmetry between $\omega_A$ and $\omega_B$.
Similarly for the case of $(i=1,j=3)$, since
$\langle\varepsilon_3(s)|\partial_s H|\varepsilon_1(s)\rangle=0$, we
find the adiabatic condition as $t_f \gg 0$. We remark that the
adiabatic condition for $(i=2,j=3)$ is the same as eq.(\ref{ad1-1}).
Therefore, considering all of the results we conclude that in order
to satisfy the adiabatic condition the total time evolution $t_f$
must satisfy eq.(\ref{ad1-2}). In what follows we will concentrate
in weak coupling limit where $H_S(t)$ dominate $H_I$. Also we
concentrate on damping term in adiabatic master equation. The
explicit form of adiabatic master equation have been presented in
eq.(\ref{density matrix evolution no ad limit}) of section appendix.
It is very important to remark that, the last terms in each of the
equations for the elements of density matrix in eq.(\ref{density
matrix evolution no ad limit}) which are proportional to the $M(s)$
are purely due to the evolution of instantaneous energy eigenbasis.
Clearly these terms are smaller than the remaining terms by a factor
$t_f$. Therefore, for sufficiently large $t_f$, when the adiabatic
condition eq.(\ref{ad1-2}) is well satisfied, we can neglect these
terms. Consequently, in the adiabatic limit the diagonal and
off-diagonal elements are decouples. In fact, in the adiabatic limit
the dynamics of the phase coherence between energy eigenstates
completely decouples from that of the energy state populations, then
their dynamics clearly do not affect the evolution of the energy
state populations. Therefore, we will investigate the dynamics
$\emph{with}$ and $\emph{without}$ adiabatic limit separately. It is
worth to remark that, to study the dynamics without adiabatic limit
we have to keep the terms including $M(s)$ in eq.(\ref{density
matrix evolution no ad limit}).

\subsection{Solution in the adiabatic limit $(M(s)=0)$}
To solve the dynamics in the adiabatic limit we will set $M(s)=0$ in
eq.(\ref{density matrix evolution no ad limit}) and find
eq.(\ref{density matrix evolution with ad limit}). Therefore, in the
adiabatic limit the dynamics of the population of the instantaneous
energy eigenstate, ($\rho_{ii}, i=1,2,3$), eq.(\ref{density matrix
evolution with ad limit}), is just because of non-zero $(x_i,
i=1,2,...10)$, as presented in eq.(\ref{xis}), which in turn
requires a non-zero $\gamma_{\alpha\beta}(\pm\Delta)$,
$\gamma_{zz}(\pm2\Delta)$, $\alpha,\beta=\{x,z\}$, i.e., a resonant
thermal excitation and relaxation. Note that, according to KMS
condition eq.(\ref{KMS}), we have:
$\gamma_{\alpha\beta}(-\Delta)=e^{-\beta
\Delta}\gamma_{\beta\alpha}(\Delta)$ and
$\gamma_{zz}(-2\Delta)=e^{-2\beta
\Delta}\gamma_{\beta\alpha}(2\Delta)$.
\subsection{Solution without adiabatic limit $(M(s)\neq0)$}
 Assuming that the bath is in a thermal state with an ohmic spectral density, \cite{Albash-2012}, we have:
\begin{equation}\label{spectral density}
\gamma_{\alpha\beta}(\omega)=2\pi \eta g^2\frac{\omega e^{-|\omega|/\omega_c}}{1-e^{-\beta \omega}},
\end {equation}

note that, $\eta$ is a positive constant with dimensions of time squared arising in the specification
of the Ohmic spectral function and $\omega_c$ is a high-frequency cutoff. We assume $g^x_k=g^z_k=g$, and plot the spectral density in Fig.1 in terms of the typical parameters used in our numerical calculations.
\begin{figure}
\centering
\includegraphics[keepaspectratio, width=1\textwidth]{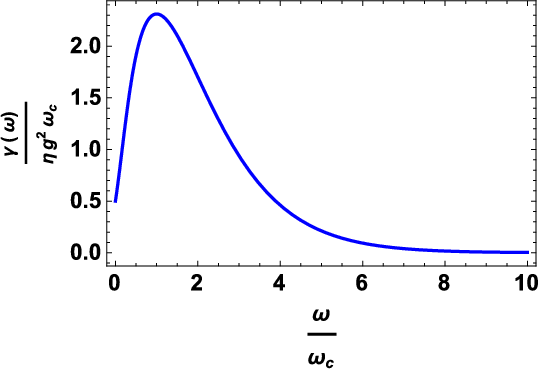}
\caption{(Color online) the Ohmic spectral density $\gamma(\omega)=2\pi \eta g^2\frac{\omega e^{-|\omega|/\omega_c}}{1-e^{-\beta \omega}}$ is depicted in terms of $\omega/\omega_c$ with $\omega_c=8 \pi~ GHz$ and $\beta^{-1}=2.6~ GHz $.}
\end{figure}
We remark that, when $\omega_c\gg 1/\beta$, the bath correlation time,
eq.(\ref{bath correlation time}), can be shown to be $\tau_B= \beta/2\pi$ \cite{Albash-2012}. Therefor, in addition to eq.(\ref{ad1-2}), the condition of eq.(\ref{Adiabatic condition}) requires that
\begin{equation}\label{openadiabatic}
t_f\gg \frac{1}{\sqrt{2}}(\frac{\beta}{2\pi})^2\sqrt{\omega_A^2+\omega_B^2}.
\end{equation}
Note that, the parameters must be chosen such that satisfy both of the inequalities.

Since our aim is to adiabatically charge three-level open QB via dark state, clearly the success probability of our charging protocol of QB highly depends on the dynamics of dark state population. Therefore, in the following by setting $\{\rho_{22}(0)=1$, and $\rho_{ij}(0)=0$ for$~ i,j\neq2 \}$, we initialize the system in the dark state $|\varepsilon_2(t)\rangle$ and examine the dependence of dark state population on total evolution time $t_f$ using eq.(\ref{density matrix evolution no ad limit}). To this end, in fig.$2$ we plot $\rho_{22}(t_f)$ as a function of $t_f$, where we explicitly see the effect of the various time scales in our problem. For very short evolution times (where the adiabatic condition is not satisfied, i.e.,$\frac{{\sqrt{2}~\omega_A^2\omega_B^2} }{(\omega_A^2+\omega_B^2)^{\frac{3}{2}}}t_f \ll 1$), the evolution is highly non-adiabatic, and the final dark state probability is close to $0$. However, note that in this regime we may not entirely trust the master equation to be a reliable approximation for the dynamics
since the condition $h/t_f\ll \tau_B^{-2}$ eq.(\ref{Adiabatic condition}) requirs that the system evolves much more slowly than the time scale of
the bath is not necessarily satisfied. Furthermore, since the evolution is so short, thermal effects are very small because they don't have enough time to act the system. Fig.2 provides a useful insight about the optimal value of evolution time $t_f$ which is also satisfies the adiabatic condition eq.(\ref{ad1-2}). Indeed, as we increase $t_f$, the evolution becomes more and more adiabatic and the dark state probability increases and arrives  close to its max value i.e., around $\frac{{\sqrt{2}~\omega_A^2\omega_B^2} }{(\omega_A^2+\omega_B^2)^{\frac{3}{2}}}t_f = 4.965$, where by setting $\omega_A=\omega_B=1$ we find $t_f = 9.93$. However, as we continue to increase $t_f$, rather than observing that the system remains in the dark state, remarkable part of its population have been lost due to increase in thermal excitations, where it is clearly can be seen in fig.2 for large $t_f$. It is worth to mention that, the optimal value of $t_f$ balances the adiabaticity of the evolution against the time allowed for thermal processes to occur.
\begin{figure}
\centering
\includegraphics[keepaspectratio, width=1\textwidth]{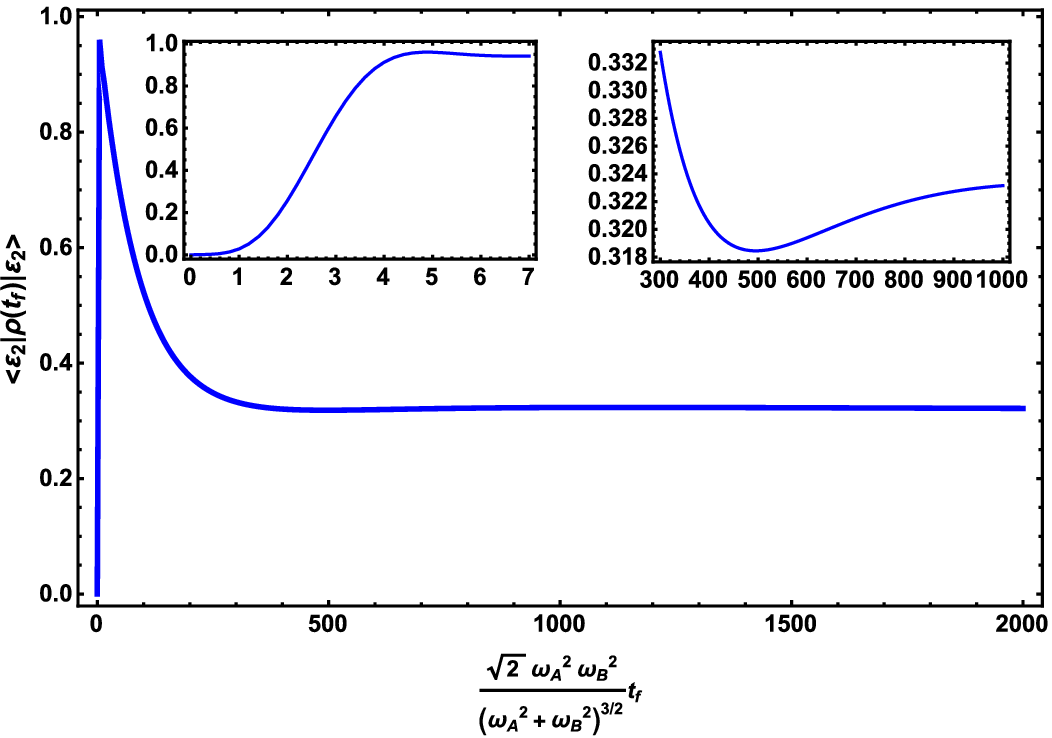}
\caption{(Color online) Final dark state population, $\rho_{22}(t_f)$, as a function of total evolution time [in dimensionless
units corresponding to the adiabatic condition eq.(\ref{ad1-2})] for the model which is governed with adiabatic master equation eq.(\ref{density matrix evolution no ad limit}) with initial condition as $\rho_{22}(0)=1$ and the parameter values as $\omega_A=\omega_B=1,~ \eta g^2=10^{-4},~ \beta=\frac{1}{2.6}$. The left inset zooms in on the short total-time evolution, the maximum in the dark state probability determines the optimal value of the evolution time $t_f$ and is seen to occur at $\frac{{\sqrt{2}~\omega_A^2\omega_B^2} }{(\omega_A^2+\omega_B^2)^{\frac{3}{2}}}t_f = 4.965$, and clarifies increasing $t_f$ much above the heuristic adiabatic condition could not help. The reason is that this maximum is a balance between maximizing adiabaticity while minimizing thermal excitations. On the other hand, the right inset shows that for long total evolution times, in the interval almost between $500-1000$, the dark state recovers part of its population due to thermal relaxation.  However, for very long evolution times $t_f$, fig.2 shows that the dark state finally settles on its Gibbs distribution value because of thermal excitation.}
\end{figure}

Here it may be a right place to ask what the impact of the parameters $\omega_{A(B)}$ of the Hamiltonian $H_S(t)$ on the optimal value of total evolution time $t_f$ is. To this aim, we will consider the short total-time evolution with different values of $\omega_{A(B)}$. Interestingly we find that the optimal value of $t_f$ highly depends on these parameters. In fact, choosing the proper values for these parameters, makes it possible to decrease suitably the optimal value of total time evolution $t_f$ and hence speed up the charging process of open QBs. As an illustration of the idea, simply we have considered different binaries as ($\omega_A=\omega_B=1$), ($\omega_A=2, \omega_B=1$), ($\omega_A=1, \omega_B=2$) and determined  the corresponding optimal values of the total evolution time, $t_f^{opt}$s. We use the same initial condition as in fig.$2$ and set the parameters as: $\eta g^2=10^{-4}, 1/\beta = 2.6 $. Fig.4 shows these results clearly, the up-left inset (Rigid green color curve) determines that for ($\omega_A=\omega_B=1$) the max value of dark state population is $0.99$ which appears at $t_f^{opt}=9.93$. The up-right inset (dashed red color curve) determines that for ($\omega_A=1, \omega_B=2$) the max value of dark state population is $0.975$ which appears at $t_f^{opt}=20.3137$, and the below inset (dotted blue curve) correspond to ($\omega_A=2, \omega_B=1$) with dark state population $0.972$ at $t_f^{opt}=20.2742$. Therefore, the proper choice of the parameter $\omega_{A(B)}$ could be very useful for improving the dark state population and speeding up the populating of dark state which in turn improves the charging process of QB.
\begin{figure}
\centering
\includegraphics[keepaspectratio, width=1\textwidth]{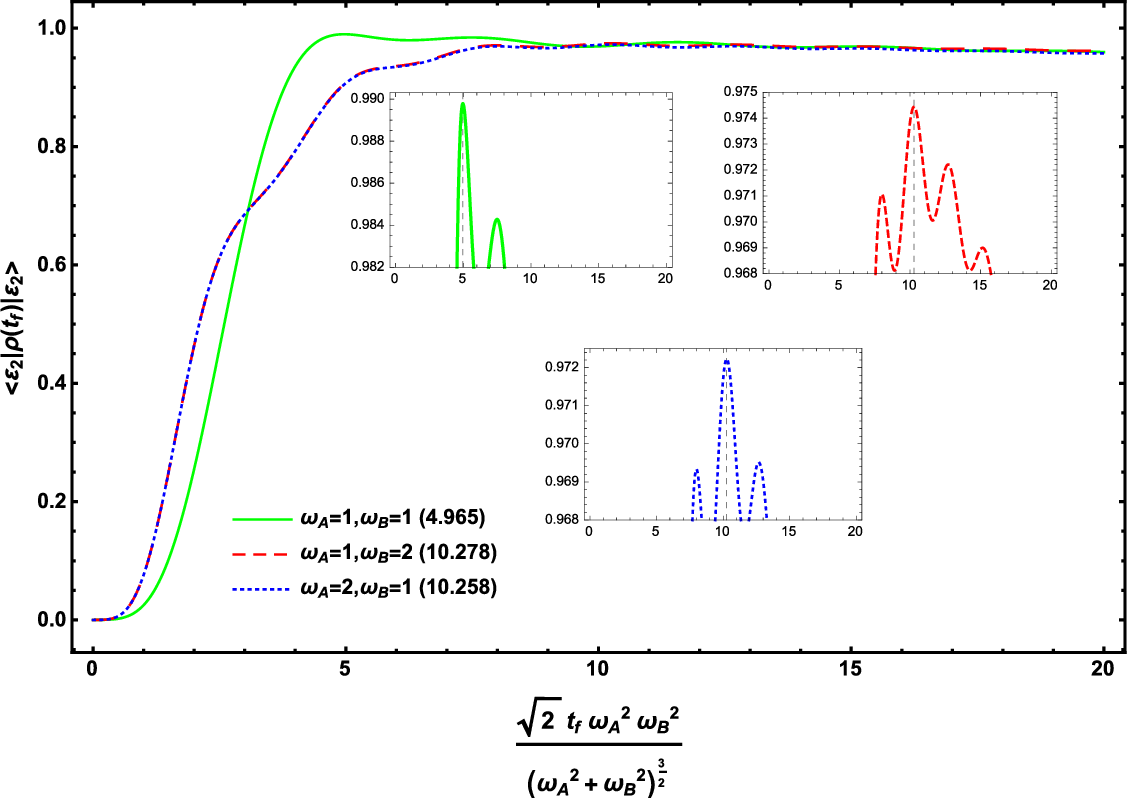}
\caption{(Color online) The effect of $\omega_{A(B)}$ on the optimal value of the evolution time.}
\end{figure}

It is also worth to recall that, in fig.2  we have shown that the final dark state population for very long evolution times $t_f$, eventually settles on its Gibbs distribution value due to thermal excitation. An interesting question would be what is distance between system density matrix and Gibss state at $t_f^{opt}$? To answer this, we computed the trace-norm distance defined in equation (\ref{trace norm distance}) between the evolving system density matrix and the instantaneous Gibss state ($\rho_{Gibss}(t)=e^{-\beta H_S(t)}/Tr[e^{-\beta H_S(t}]$) at $t_f^{opt}$. In fig.$4$, we determine the trace-norm distance between the system density matrix and the Gibss state for different set of values chosen for the parameters of system as ($\omega_A=\omega_B=1,~t_f^{opt}=9.93$, rigid green curve ), ($\omega_A=2, \omega_B=1$, $t_f^{opt}=20.2742$, blue dotted curve), and ($\omega_A=1, \omega_B=2, t_f^{opt}=20.3137$, red dashed curve ) with $\eta g^2=10^{-4}, 1/\beta = 2.6$. The results show that for all of the above examples with different $t_f^{opt}$ the
system is far enough from its corresponding Gibss state.
\begin{figure}
\centering
\includegraphics[keepaspectratio, width=1\textwidth]{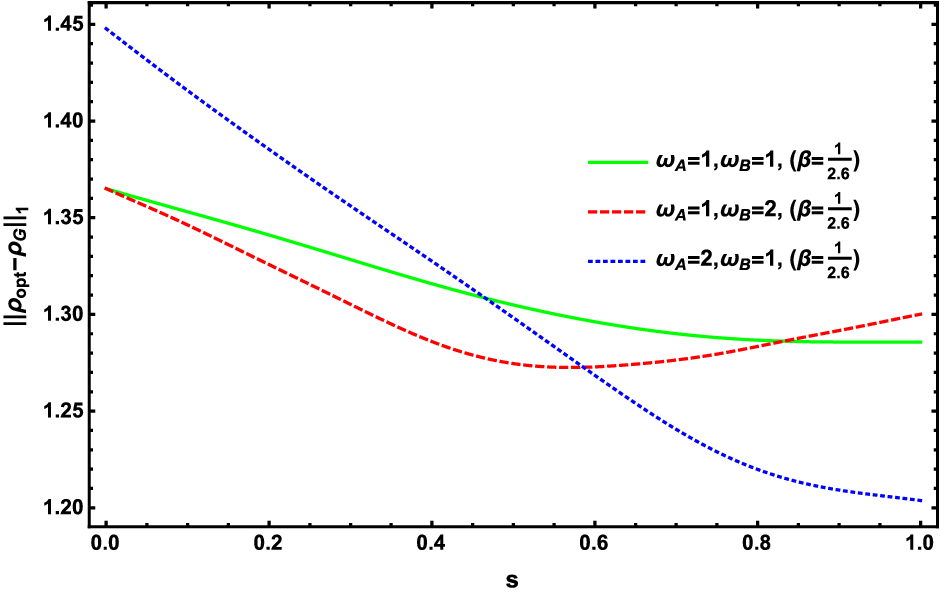}
\caption{(Color online) Trace-norm distance between the evolving system density matrix
(using the Lindblad adiabatic master equation eq.(\ref{density matrix evolution no ad limit})) and the Gibbs state at $t_f^{opt}$.}
\end{figure}

Furthermore, in the following, in fig.$5$, we describe the dependence of optimal total time evolution on the bath strength. One can see that as the system-bath strength increases, such that thermal processes occur more rapidly, the optimal evolution time decreases. This presents a remarkable feature of dynamics of open quantum system via adiabatic master equation. Indeed, in the charging of an open QB based on adiabatic master equation we can stop the charging early before the dark state probability starts to fall down due to thermal excitations, which in turn provides a potential advantageous of speeding up in charging process of QBs with this protocol.
\begin{figure}
\centering
\includegraphics[keepaspectratio, width=1\textwidth]{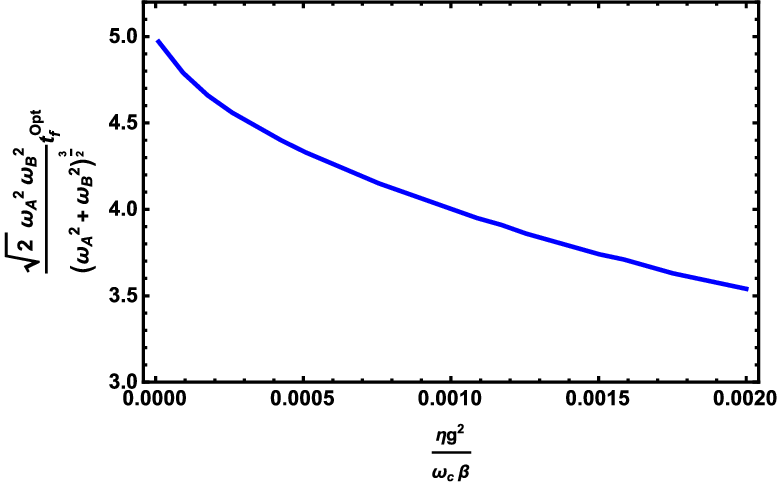}
\caption{(Color online) Trace-norm distance between the evolving system density matrix
(using the Lindblad adiabatic master equation eq.(\ref{density matrix evolution no ad limit})) and the Gibbs state at $t_f^{opt}$.}
\end{figure}

Based on these considerations, now we have all ingredient and enough insight and knowledge to investigate adiabatic charging of open three-level QB. To this end, we study the dynamics of adiabatic charging process, including dynamics of stored energy, ergotropy, and efficiency for different total evolution time $t_f=\{0.1,~9.93,~ 500,~ 2000\}$. The results are seen in fig.$6$, we show that for very small total time evolution $t_f=0.1$, where the heuristic adiabatic condition is not satisfied, the stored energy $\Delta E$, ergotropy $W$, are very small as it is expected from our results obtained in dynamics of dark-state population. We show that at optimal value of $t_f^{Opt}=9.93$, the battery get fully charged and the ergotropy achieves its maximum value and the efficiency $\eta$ becomes equal to one. In fact, with charging time equal to optimal value of $t_f=9.93$ we can extract the total maximum stored energy $\Delta E= 1.95$ from battery. Note that we set the eigen values of the bare Hamiltonian $H_0$ as $\lambda_1=0 $, $\lambda_2=\hbar \omega$, and $\lambda_3=1.95\hbar \omega$. As we continue to increase $t_f$ from its optimal value, to $t_f=500$, obviously we find that the stored energy $\Delta E$, ergotropy $W$, as well as the efficiency $\eta$ start to decrease from their maximum values because of thermal excitation. It is clearly seen in fig.$6$, that increasing the total evolution time more by setting $t_f=2000$, where the adiabatic condition well satisfied, the situation for the stored energy and ergotropy get more worse. Consequently our investigation reveals that increasing $t_f$ much more than its optimal value not only dose not help to the charging process but also is harmful for it.
\begin{figure}
\centering
\includegraphics[keepaspectratio, width=1\textwidth]{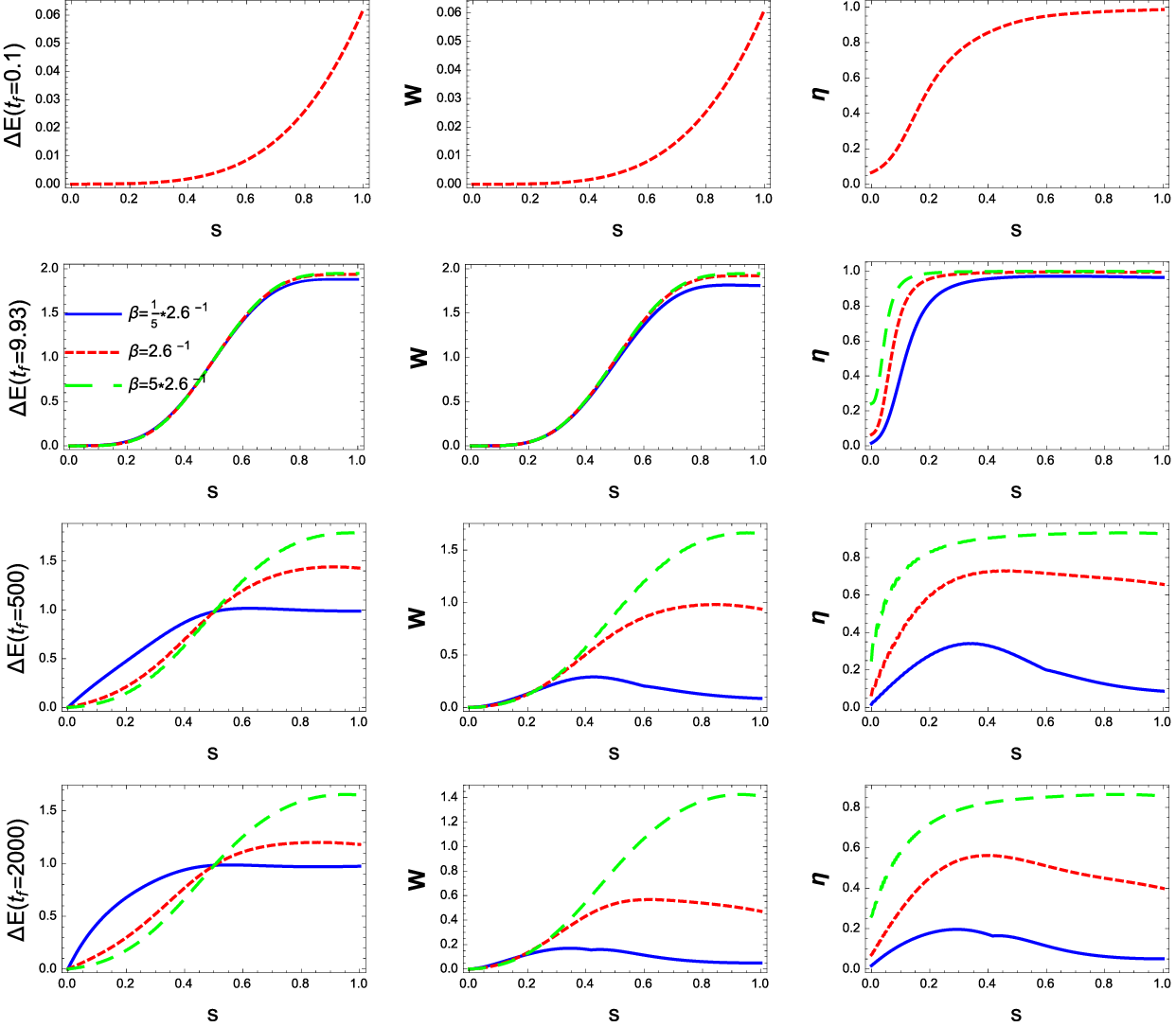}
\caption{(Color online) Dynamics of stored energy $\Delta E$, ergotropy $W$, and efficiency $\eta$, as a function of parameter $s$ for different scales of evolution time $t_f$ have been depicted according to adiabatic master equation eq.(\ref{density matrix evolution no ad limit}) with the same initial condition as specified in fig.$2$. As it is seen, for total time evolution with optimal value of $t_f=9.93$ the battery is fully charged and the total energy of battery can be extracted completely which leads to $\eta=1$. For $t_f=9.93$ the heuristic adiabatic condition is satisfied. However for $t_f$s with scale much smaller than the optimal $t_f$, where the adiabatic limit is not satisfied, the stored energy $\Delta E$, ergotropy $W$, and efficiency $\eta$ are very small(we depict the results just for $\beta=2.6^{-1}$  ). Moreover for time evolution much larger than optimal $t_f$ although the adiabatic limit is satisfied, the stored energy $\Delta E$, ergotropy $W$, and efficiency $\eta$ are smaller in comparison with those results of optimal time. Meanwhile, for a given $t_f$, we analyze the effect of environment temperature $1/\beta$, on stored energy $\Delta E$, ergotrop $W$, and efficiency $\eta$. The results show that as we increase the temperature of environment, $\Delta E$, $W$, and efficiency $\eta$ decrease and vise versa.}
\end{figure}
In addition, in order to examine the effect of environment temperature on charging process, in fig.$6$, with a given $t_f$ we have depict $\Delta E$, $W$, and efficiency $\eta$ for different values of environment temperature $\beta=\{1/5(2.6^{-1}),2.6^{-1},5(2.6^{-1})\}$ (note that the environment temperature equals to inverse of $\beta$). As an example, considering for $t_f=9.93$, the results show that at low temperatures $\beta=5(2.6^{-1})$, rigid green curve, battery is almost fully charged ($\Delta E=1.95$), as we increase the temperature by setting $\beta=2.6^{-1}$, dashed red curve, the stored energy of battery decreases. If we increase temperature even more by setting $\beta=1/5(2.6^{-1})$, the stored energy of battery decrease even more, as it is seen in blue rigid curve. We have the same results for ergotropy and efficiency in terms of temperature as those of stored energy. Fig.$6$ shows that $\Delta E$, $W$, and efficiency $\eta$ have the same behavior for other $t_f$s in terms of different $\beta$s as those for $t_f=9.93$.

Therefore, considering fig.$6$ and comparison of stored energy $\Delta E$, ergotropy $W$ and efficiency $\eta$ for a given $t_f$ with different values of temperature of environment reveals that by increasing the temperature of environment, stored energy, ergotropy and efficiency of QB decrease and vise versa. These results can be understood by concerning that, as the temperature of environment increases, it could help the process of thermal excitation and consequently the population of dark state decreases more.
Thus, with higher temperature of environment, thermal excitations can have a significant detrimental impact on the dark state population, and hence destructively acts on the success probability of an adiabatic quantum charging process. Therefore, higher the temperature of environment is the lower stored energy, ergotropy $W$, as well as the efficiency $\eta$ of QB.

To complete our discussion in the following lines we investigate the effect of Hamiltonian parameters $\omega_A$ and $\omega_B$ on $\Delta E$, ergotropy $W$ and efficiency $\eta$. To this aim and as an example to present the idea simply, we consider different set of values for the parameters of system as ($\omega_A=\omega_B=1,~t_f^{opt}=9.93$, rigid green curve ), ($\omega_A=2, \omega_B=1$, $t_f^{opt}=20.2742$, blue dotted curve), and ($\omega_A=1, \omega_B=2, t_f^{opt}=20.3137$, red dashed curve ) and determine $\Delta E$, ergotropy $W$ and efficiency $\eta$ for each of these cases. Note that we set $\eta g^2=10^{-4}$ and $\beta = 2.6^{-1}$. Fig.$7$ presents the results, as it is seen for the case of ($\omega_A=\omega_B=1$), the battery is fully charged, that is, $\Delta E=1.95$ at $t_f^{opt}=9.93$. Meanwhile, we could extract the total stored energy from battery at $t_f^{opt}=9.93$ ($W=1.95$), and the efficiency is equal to $1$ at $t_f^{opt}=9.93$. However, for the other two cases ($\omega_A=2(1), \omega_B=1(2))$ not only dose the charging process take long time as $(t_f^{opt}=20.2742( t_f^{opt}=20.3137))$ but also the battery is not fully charged at corresponding optimal time evolution. Note that, the same conclusion is true for ergotropy and efficiency. Therefore, we conclude that proper choice of the system parameters values are very important and could effectively guarantees the success probability of charging process of open QBs.
\begin{figure}
\centering
\includegraphics[keepaspectratio, width=1\textwidth]{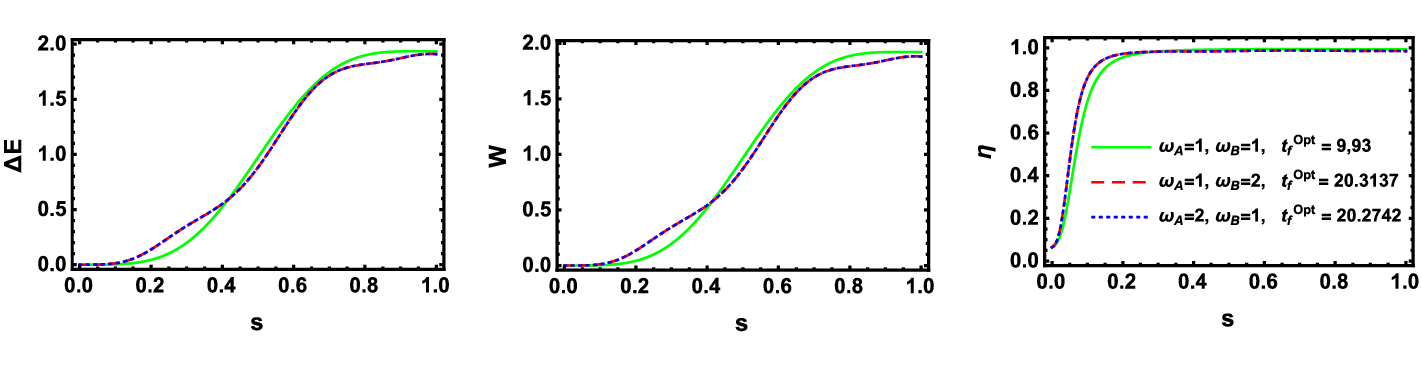}
\caption{(Color online) Dynamics of stored energy, ergotropy and efficiency of QB for different binary set of $(\omega_A, \omega_B)$ with corresponding $t_f^{opt}$s.}
\end{figure}
\section{Conclusion}
In this work, we reconsider the problem of stable adiabatic charging of open 3-level QBs, using adiabatic master equation
approach and dark state. We restrict our discussion to the weak coupling limit. In particular, a remarkable result we have shown here
(see fig. $2$) that there exists an optimal evolution time $t_f$ (problem-dependent) that is much shorter than the adiabatic time
scale, where the dark state population can be significantly higher than that of the thermal Gibbs state which in turns lead to the full charging of QB with high efficiency which is almost equal to one. Meanwhile, we have demonstrated that this optimal $t_f$ depends on system parameters (see fig. $3$) and decreases with the strength of the system-bath coupling (see fig. $5$). This shows that it can be advantageous to charge QB with a much shorter duration total evolution than that suggested by the standard (heuristic) inverse gap criterion. Indeed, these result provide an interesting technique to speedup the charging process (see fig. $6$).
Moreover, we have investigated the impact of environment temperature on charging process and shown that increasing the temperature give rise to the decrease in stored energy,  ergotropy and efficiency of QBs and vise versa (see fig. $7$).
It is also worth to remark that according to KMS condition $\gamma_{\alpha\beta}(-\Delta)=e^{-\beta \Delta}\gamma_{\beta\alpha}(\Delta)$, increasing the systems energy gap could be the standard strategy of suppressing detrimental thermal excitations.

Indeed, our protocol can be implemented with superconducting circuit quantum electrodynamics system and trapped ion systems \cite{Devoret-2013,Wendin-2017,Gu-2017}. In particular superconducting transmon qubits would be a suitable candidates for ladder-type
three-level system \cite{Koch-2007,You-2007,Barends-2013} as proposed in \cite{Santos-2019} and experimentally verified by \cite{Santos-2022}.
Although the main idea of our protocol almost is the same as that of \cite{Santos-2019}, there is a remarkable difference between solutions of our approach and those of mentioned works because of exploiting of Markovian adiabatic master equation instead of phenomenological considerations. Indeed, Investigation of adiabatic charging of QB based on Markovian adiabatic master equation clarifies that; in the open-system setting, contrary to the closed-system, where the only relevant time scale is $T_{ad}$, there are different relevant time scales, in addition to $T_{ad}$, determining the efficiency of the charging process including the time evolution $t_f$, the relaxation time $T_1$, the dephasing time $T_2$. The interplay between these time scales is nonmonotonic and certainly more complicated than in the closed-system setting. The proposed protocol not only improve our theoretical knowledge in stable adiabatic charging of open QB but also could provide new steps toward its experimental realization. This protocol opens the way for further theoretical investigation in adiabatic quantum computing and improve our understanding about effective and fast charging of open QBs.
\section{Appendix: Lindblad operators and Adiabatic master equation}

We introduce the explicit form of the time dependent Lindblad operators, eq.(\ref{lindblad operators}), in weak coupling limit, as
\begin{eqnarray}\label{lindblad operators1}
\begin{cases}
\begin{array}{ccc}
L^{\dagger}_{(x,0)}=L_{(x,0)}=\frac{A(s)+B(s)}{\sqrt{2}\Delta(s)}\big(-|\varepsilon_1\rangle\langle\varepsilon_1|
+|\varepsilon_3\rangle\langle\varepsilon_3\big)\\
L^{\dagger}_{(x,-\Delta(s))}=L_{(x,\Delta(s))}=-\frac{A(s)-B(s)}{\Delta(s)}\big(|\varepsilon_1\rangle\langle\varepsilon_2|
-|\varepsilon_2\rangle\langle\varepsilon_3\big)\\
L^{\dagger}_{(x,\Delta(s))}=L_{(x,-\Delta(s))}=-\frac{A(s)-B(s)}{\Delta(s)}\big(|\varepsilon_2\rangle\langle\varepsilon_1|
-|\varepsilon_3\rangle\langle\varepsilon_2\big)\\
L^{\dagger}_{(x,-2\Delta(s))}=L_{(x,2\Delta(s))}=0\\
L^{\dagger}_{(x,2\Delta(s))}=L_{(x,-2\Delta(s))}=0\\
L^{\dagger}_{(z,0)}=L_{(z,0)}=\frac{A^2(s)-B^2(s)}{2\Delta^2(s)}\big(|\varepsilon_1\rangle\langle\varepsilon_1|
-2|\varepsilon_2\rangle\langle\varepsilon_2+|\varepsilon_3\rangle\langle\varepsilon_3\big)\\
L^{\dagger}_{(z,-\Delta(s))}=L_{(z,\Delta(s))}=\frac{-\sqrt{2}A(s)B(s)}{\Delta^2(s)}\big(|\varepsilon_1\rangle\langle\varepsilon_2|
+|\varepsilon_2\rangle\langle\varepsilon_3\big)\\
L^{\dagger}_{(z,\Delta(s))}=L_{(z,-\Delta(s))}=\frac{-\sqrt{2}A(s)B(s)}{\Delta^2(s)}\big(|\varepsilon_2\rangle\langle\varepsilon_1|
+|\varepsilon_3\rangle\langle\varepsilon_2\big)\\
L^{\dagger}_{(z,-2\Delta(s))}=L_{(z,2\Delta(s))}=\frac{A^2(s)-B^2(s)}{2\Delta^2(s)}|\varepsilon_1\rangle\langle\varepsilon_3|
\\
L^{\dagger}_{(z,2\Delta(s))}=L_{(z,-2\Delta(s))}=\frac{A^2(s)-B^2(s)}{2\Delta^2(s)}|\varepsilon_3\rangle\langle\varepsilon_1|
\\
\end{array}
\end{cases}
\end{eqnarray}
Plugging in these to the adiabatic master equation eq.(\ref{master}) and denoting $\rho_{ij}=\langle\varepsilon_i(s)|\rho(s)|\varepsilon_j(s)\rangle$ and using the below identities,

\begin{eqnarray}\label{eigenvectortime derivitive}
\begin{cases}
\begin{array}{ccc}
\langle\varepsilon_2(s)|\partial_s|\varepsilon_1(s)\rangle=-\langle\varepsilon_1(s)|\partial_s|\varepsilon_2(s)\rangle=\frac{A(s)\partial_s B(s)-B(s)\partial_s A(s)}{\sqrt{2}\Delta(s)^2}=M(s),\\
\langle\varepsilon_3(s)|\partial_s|\varepsilon_1(s)\rangle=\langle\varepsilon_1(s)|\partial_s|\varepsilon_3(s)\rangle=(\partial_s\langle\varepsilon_1(s)|)|\varepsilon_3(s)\rangle=
(\partial_s\langle\varepsilon_3(s)|)|\varepsilon_1(s)\rangle=0,\\
\langle\varepsilon_2(s)|\partial_s|\varepsilon_3(s)\rangle=-\langle\varepsilon_3(s)|\partial_s|\varepsilon_2(s)\rangle=M(s),\\
\langle\varepsilon_i(s)|\partial_s|\varepsilon_i(s)\rangle=(\partial_s\langle\varepsilon_i(s)|)|\varepsilon_i(s)\rangle=0,~~~i=1,2,3,\\
(\partial_s\langle\varepsilon_1(s)|)|\varepsilon_2(s)\rangle=-(\partial_s\langle\varepsilon_2(s)|)|\varepsilon_1(s)\rangle=M(s),\\
(\partial_s\langle\varepsilon_3(s)|)|\varepsilon_2(s)\rangle=-(\partial_s\langle\varepsilon_2(s)|)|\varepsilon_3(s)\rangle=M(s),\\
\end{array}
\end{cases}
\end{eqnarray}

we achieve the following equations for the elements of the Hermitian density matrix in the time dependent instantaneous energy eigenbasis
\begin{eqnarray}\label{density matrix evolution no ad limit}
\hspace{0mm}\begin{cases}
   \begin{array}{cc}

\dot{\rho_{11}} = (\sum_{i=1}^{4}{x_i})t_f\rho_{22}-(\sum_{i=5}^{9}{x_i})t_f\rho_{11}+x_{10}t_f\rho_{33}+M(s)(\rho_{12}+\rho_{21}),\\

\dot{\rho_{12}} = \bigg(-\frac{1}{2} y_1+\frac{5}{2}y_3+\frac{1}{2}y_4-\frac{9}{2}y_2-\frac{1}{2}\sum_{i=1}^{4}{x_i}-x_5-x_8
-\frac{1}{2}x_9\bigg)t_f \rho_{12}+\\(-x_1+x_2-x_3+x_4)t_f \rho_{23}-M(s)(\rho_{11}+\rho_{13}-\rho_{22}),\\

\dot{\rho_{13}} = \bigg(-2y_1-
y_3+y_4-\frac{1}{2}x_1+\frac{1}{2}x_2+\frac{1}{2}x_3-\frac{1}{2}\sum_{i=4}^{10}{x_i}\bigg)t_f \rho_{13}+\\M(s)(\rho_{12}+\rho_{23}),\\

\dot{\rho_{21}} = \bigg(-\frac{1}{2} y_1+\frac{1}{2}y_3+\frac{5}{2}y_4-\frac{9}{2}y_2-\frac{1}{2}\sum_{i=1}^{4}{x_i}-x_5-x_8
-\frac{1}{2}x_9\bigg)t_f \rho_{21}+\\(-x_1-x_2+x_3+x_4)t_f\rho_{32}-M(s)(\rho_{11}+\rho_{31}-\rho_{22}),\\\\

\dot{\rho_{22}} = (\sum_{i=1}^{5}{x_i}+x_6+x_7-x_8)t_f\rho_{22}+\bigg(x_1-x_2
-x_3+x_4\bigg)t_f\rho_{33}+(\sum_{i=5}^{8}{x_i})t_f\rho_{11}-\\M(s)(\rho_{12}+\rho_{21}+\rho_{23}+\rho_{32}),\\\\

\dot{\rho_{23}} = \bigg(-\frac{1}{2}y_1-\frac{1}{2}y_3-\frac{5}{2}y_4-\frac{9}{2}y_2-x_1-x_4
-\frac{1}{2}x_5+\frac{1}{2}x_6+\\\frac{1}{2}x_7-\frac{1}{2}x_8-\frac{1}{2}x_{10}\bigg)t_f \rho_{23}+(-x_5+x_6
-x_7+x_8)t_f \rho_{12}+M(s)(\rho_{22}-\rho_{13}-\rho_{33}),\\\\

\dot{\rho_{31}} = \bigg(-2y_1+
y_3-y_4-\frac{1}{2}x_1+\frac{1}{2}x_2+\frac{1}{2}x_3-\frac{1}{2}\sum_{i=4}^{10}{x_i}\bigg)\rho_{31}+\\M(s)(\rho_{21}+\rho_{32}),\\

\dot{\rho_{32}} = \bigg(-\frac{1}{2}y_1-\frac{5}{2}y_3-\frac{1}{2}y_4-\frac{9}{2}y_2-x_1-x_4
-\frac{1}{2}x_5+\frac{1}{2}x_6+\\\frac{1}{2}x_7-\frac{1}{2}x_8-\frac{1}{2}x_{10}\bigg)t_f \rho_{32}+(-x_5+x_6
-x_7+x_8)t_f \rho_{21}+M(s)(\rho_{22}-\rho_{31}-\rho_{33}),\\

\dot{\rho_{33}} =(-x_1+x_2+x_3-x_4-x_{10})t_f\rho_{33}+(x_5-x_6-x_7+x_8)t_f\rho_{22}+
x_9t_f\rho_{11}\\+M(s)(\rho_{23}+\rho_{32}).\\
\end{array}
\end{cases}
\end{eqnarray}
We remark that the derivatives are with respect to $s$ $(\partial_t=\frac{1}{t_f}\partial_s)$ and the time dependent
parameters $x_i$ and $y_i$ are as follow
\begin{eqnarray}\label{xis}
\begin{cases}
\begin{array}{ccc}
x_{1}=\frac{(A(s)-B(s))^2\gamma_{xx}(\Delta(s))}{4\Delta(s)^2},~~~
x_{2}=\frac{(A(s)-B(s))^2\gamma_{xx}(-\Delta(s))}{4\Delta(s)^2},\\
x_{3}=\frac{A(s)B(s)(A(s)-B(s))\gamma_{xz}(\Delta(s))}{\sqrt{2}\Delta(s)^3},~~~
x_{4}=\frac{A(s)B(s)(A(s)-B(s))\gamma_{xz}(-\Delta(s))}{\sqrt{2}\Delta(s)^3},\\
x_{5}=\frac{A(s)B(s)(A(s)-B(s))\gamma_{zx}(\Delta(s))}{\sqrt{2}\Delta(s)^3},~~~
x_{6}=\frac{A(s)B(s)(A(s)-B(s))\gamma_{zx}(-\Delta(s))}{\sqrt{2}\Delta(s)^3},\\
x_{7}=\frac{2A(s)^2B(s)^2\gamma_{zz}(\Delta(s))}{\Delta(s)^4},~~~
x_{8}=\frac{2A(s)^2B(s)^2\gamma_{zz}(-\Delta(s))}{\Delta(s)^4},\\
x_{9}=\frac{(A(s)^2-B(s)^2)^2\gamma_{zz}(2\Delta(s))}{4\Delta(s)^4},~~~
x_{10}=\frac{(A(s)^2-B(s)^2)^2\gamma_{zz}(-2\Delta(s))}{4\Delta(s)^4},\\
y_{1}=\frac{(A(s)+B(s))^2\gamma_{xx}(0)}{2\Delta(s)^2},~~~
y_{2}=\frac{A(s)^2-B(s)^2)^2\gamma_{zz}(0)}{4\Delta(s)^4}\\
y_{3}=\frac{(A(s)+B(s))(A(s)^2-B(s)^2)\gamma_{xz}(0)}{2\sqrt{2}\Delta(s)^3},~~~
y_{4}=\frac{(A(s)+B(s))(A(s)^2-B(s)^2)\gamma_{zx}(0)}{2\sqrt{2}\Delta(s)^3}.\\
\end{array}
\end{cases}
\end{eqnarray}
We emphasize that, the last terms in the above set of differential equations, eq.(\ref{density matrix evolution no ad limit}), which is proportional to the $M(s)$, is purely because of the evolution of instantaneous energy eigenbasis. On the other hand, in each differential equation these terms are smaller than the remaining terms with a factor $t_f$. Therefore, for sufficiently large $t_f$, when the adiabatic condition eq.(\ref{ad1-1}) is satisfied, we can neglect these terms. Consequently, in the adiabatic limit we set $M(s)=0$, and as a result the diagonal and off-diagonal elements are decouples as

\begin{eqnarray}\label{density matrix evolution with ad limit}
\hspace{0mm}\begin{cases}
   \begin{array}{cc}

\dot{\rho_{11}} = (\sum_{i=1}^{4}{x_i})t_f\rho_{22}-(\sum_{i=5}^{9}{x_i})t_f\rho_{11}+x_{10}t_f\rho_{33},\\
\dot{\rho_{22}} = (\sum_{i=1}^{5}{x_i}+x_6+x_7-x_8)t_f\rho_{22}+\bigg(x_1-x_2
-x_3+x_4\bigg)t_f\rho_{33}+(\sum_{i=5}^{8}{x_i})t_f\rho_{11},\\
\dot{\rho_{33}} =(-x_1+x_2+x_3-x_4-x_{10})t_f\rho_{33}+(x_5-x_6-x_7+x_8)t_f\rho_{22}+
x_9t_f\rho_{11}.\\\\

\dot{\rho_{12}} = \bigg(-\frac{1}{2} y_1+\frac{5}{2}y_3+\frac{1}{2}y_4-\frac{9}{2}y_2-\frac{1}{2}\sum_{i=1}^{4}{x_i}-x_5-x_8
-\frac{1}{2}x_9\bigg)t_f \rho_{12}+\\(-x_1+x_2-x_3+x_4)t_f \rho_{23},\\

\dot{\rho_{13}} = \bigg(-2y_1-
y_3+y_4-\frac{1}{2}x_1+\frac{1}{2}x_2+\frac{1}{2}x_3-\frac{1}{2}\sum_{i=4}^{10}{x_i}\bigg)t_f \rho_{13},\\

\dot{\rho_{21}} = \bigg(-\frac{1}{2} y_1+\frac{1}{2}y_3+\frac{5}{2}y_4-\frac{9}{2}y_2-\frac{1}{2}\sum_{i=1}^{4}{x_i}-x_5-x_8
-\frac{1}{2}x_9\bigg)t_f \rho_{21}+\\(-x_1-x_2+x_3+x_4)t_f\rho_{32},\\

\dot{\rho_{23}} = \bigg(-\frac{1}{2}y_1-\frac{1}{2}y_3-\frac{5}{2}y_4-\frac{9}{2}y_2-x_1-x_4
-\frac{1}{2}x_5+\frac{1}{2}x_6+\\\frac{1}{2}x_7-\frac{1}{2}x_8-\frac{1}{2}x_{10}\bigg)t_f \rho_{23}+(-x_5+x_6
-x_7+x_8)t_f \rho_{12},\\

\dot{\rho_{31}} = \bigg(-2y_1+
y_3-y_4-\frac{1}{2}x_1+\frac{1}{2}x_2+\frac{1}{2}x_3-\frac{1}{2}\sum_{i=4}^{10}{x_i}\bigg)\rho_{31},\\

\dot{\rho_{32}} = \bigg(-\frac{1}{2}y_1-\frac{5}{2}y_3-\frac{1}{2}y_4-\frac{9}{2}y_2-x_1-x_4
-\frac{1}{2}x_5+\frac{1}{2}x_6+\\\frac{1}{2}x_7-\frac{1}{2}x_8-\frac{1}{2}x_{10}\bigg)t_f \rho_{32}+(-x_5+x_6
-x_7+x_8)t_f \rho_{21}.\\

\dot{\rho_{33}} =(-x_1+x_2+x_3-x_4-x_{10})t_f\rho_{33}+(x_5-x_6-x_7+x_8)t_f\rho_{22}+
x_9t_f\rho_{11}.\\
\end{array}
\end{cases}
\end{eqnarray}

In fact, in the adiabatic limit in which the dynamics of the phase coherence between energy eigenstates completely decouples from that of the energy state populations, such that their dynamics do not affect the evolution of the energy state populations. Therefore it is important to investigate the dynamics with and without adiabatic limit separately. It is worth to remark that to study the dynamics without adiabatic limit we have to keep the terms including $M(s)$ in eq.(\ref{density matrix evolution no ad limit})

\subsection{Trace norms distance}
The trace norm distance is an unitarily invariant norm that
satisfies, for all unitary U, V, and for any operator $A:{||UAV||}_{ui} =|A|_{ui}$. Suppose $|A|\equiv\sqrt{A^\dag A}$,  the trace norm of $||A||_{1}$ is defined as
\begin{equation}\label{trace norm distance}
||A||_{1}=Tr|A|=\sum_{i}{s_i(A)}
\end{equation}
where $s_i(A)$ are the eigenvalues of $|A|$. The unitarily invariant norms satisfy some inequalities; for more details see, e.g., \cite{Bhatia-1997,Lidar4-2008}.
\\\\
\textbf{\large{Data availability}}\\ The datasets used and analysed
during the current study available from the corresponding author on
reasonable request.


\begin{thebibliography}{99}
\bibitem{Vinjanampathy-2016} S. Vinjanampathy and J. Anders, CONTEMP PHYS 57, 545 (2016).
\bibitem{Alicki-2018} R. Alicki and R. Kosloff, Introduction to quantum thermodynamics: History and prospects, in Thermodynamics in the Quantum Regime (Springer, 2018).
\bibitem{Deffner-2019} S. Deffner and S. Campbell, Quantum Thermodynamics: An introduction to the thermodynamics of quantum information
    (Morgan and Claypool Publishers, 2019).
\bibitem{Alicki-2013} R. Alicki and M. Fannes, Phys. Rev. E 87, 042123 (2013).
\bibitem{Andolina-2018} G. M. Andolina, D. Farina, A. Mari, V. Pellegrini, V. Giovannetti, and M. Polini, Phys. Rev. B 98, 205423 (2018).
\bibitem{Giov-2019} D. Farina, G. M. Andolina, A. Mari, M. Polini, and V. Giovannetti, Charger-mediated energy transfer for quantum batteries: An open-system approach, Phys. Rev. B 99, 035421 (2019).
\bibitem{Santos-2020} A. C. Santos, A. Saguia, and M. S. Sarandy, Phys. Rev. E 101, 062114 (2020).
\bibitem{Le-2018} T. P. Le, J. Levinsen, K. Modi, M. M. Parish, and F. A. Pollock, Phys. Rev. A 97, 022106 (2018).
\bibitem{Wang-2019} Y.-Y. Zhang, T.-R. Yang, L. Fu, and X. Wang, Phys. Rev. E 99, 052106 (2019).
\bibitem{Rossini-2019} D. Rossini, G. M. Andolina, and M. Polini, Phys. Rev. B 100, 115142 (2019).
\bibitem{Kamin-2020} F. H. Kamin, F. T. Tabesh, S. Salimi, and A. C. Santos, Phys. Rev. E 102, 052109 (2020).
\bibitem{Crescente-2020} A. Crescente, M. Carrega, M. Sassetti, and D. Ferraro, Phys. Rev. B 102, 245407 (2020).
\bibitem{Munro-2020} J. Q. Quach and W. J. Munro, Using dark states to charge and stabilize open quantum batteries, Phys. Rev. Appl. 14, 024092
    (2020).
\bibitem{Tabesh-2020} F. T. Tabesh, F. H. Kamin, and S. Salimi, Phys. Rev. A 102, 052223 (2020).
\bibitem{Song-2022} M. L. Song, L. J. Li, X. K. Song, L. Ye, and D. Wang, Phys. Rev. E 106, 054107 (2022).
\bibitem{Morrone-2023} D. Morrone1, M. A. C. Rossi, A. Smirne, and M. G. Genoni, Quantum Sci. Technol. 8, 035007 (2023).
\bibitem{Xu-2024} K. Xu, H. G. Li, H. J. Zhu, and W. M. Liu, Phys. Rev. E 109, 054132 (2024).
\bibitem{Mojaveri-2024} B. Mojaveri, R. Jafarzadeh Bahrbeig, and M. A. Fasihi, Extracting ergotropy from nonequilibrium steady states of an XXZ spin-chain quantum battery, Phys. Rev. A 109, 042619 (2024).
\bibitem{Ghosh-2021} S. Ghosh, T. Chanda, S. Mal, and A. Sen, Phys. Rev. A 104, 032207 (2021).
\bibitem{Mojaveri1-2024} B. Mojaveri, R. Jafarzadeh Bahrbeig, and M. A. Fasihi, Charging a Quantum Battery Mediated by Parity-Deformed Fields,
    https://doi.org/10.48550/arXiv.2405.11356.
\bibitem{Mojaveri-2023} B. Mojaveri, R. Jafarzadeh Bahrbeig, M. A. Fasihi, and S. Babanzadeh, Enhancing the direct charging performance of an open quantum battery by adjusting its velocity, Sci. Rep. 13, 19827 (2023).
\bibitem{Mitchison-2021} M. T. Mitchison, J. Goold, and J. Prior, Charging a quantum battery with linear feedback control, Quantum 5, 500 (2021).
\bibitem{Yao-2022} Y. Yao and X. Q. Shao, Optimal charging of open spin-chain quantum batteries via homodyne-based feedback control, Phys. Rev. E 106, 014138 (2022).
\bibitem{Koshihara-2023} K. Koshihara and K. Yuasa, Quantum ergotropy and quantum feedback control, Phys. Rev. E 107, 064109 (2023).
\bibitem{Pasquale-2017} A. De Pasquale, K. Yuasa, and V. Giovannetti, Phys. Rev. A 96, 012316 (2017).
\bibitem{Gherardini-2019} S. Gherardini, A. Smirne, M. M. Muller, and F. Caruso, Proceedings 12, 11 (2019).
\bibitem{Gherardini-2020} S. Gherardini, F. Campaioli, F. Caruso, and F. C. Binder, Stabilizing open quantum batteries by sequential measurements Phys. Rev. Research 2, 013095 (2020).
\bibitem{Santos-2019} A. C. Santos, B. Cakmak, S. Campbell, and N. T. Zinner, Stable adiabatic quantum batteries, Phys. Rev. E 100,
    032107 (2019).
\bibitem{Scully-1989} M. O. Scully, S.-Y. Zhu, and A. Gavrielides, Phys. Rev. Lett. 62, 2813 (1989).
\bibitem{Mompart-2000} J. Mompart and R. Corbalan, J. Opt. B: Quantum Semiclassical Opt. 2, R7 (2000).
\bibitem{Gaubatz-1990} U. Gaubatz, P. Rudecki, S. Schiemann, and K. Bergmann, J. Chem. Phys. 92, 5363 (1990).
\bibitem{Vitanov-2017} N. V. Vitanov, A. A. Rangelov, B. W. Shore, and K. Bergmann, Rev. Mod. Phys. 89, 015006 (2017).
\bibitem{Shore-2017} B. W. Shore, Picturing stimulated Raman adiabatic passage: a STIRAP tutorial, Adv. Opt. Photon 9, 563 (2017).
\bibitem{Liu-2019} J. Liu, D. Segal, and G. Hanna, Loss-free excitonic quantum battery, J. Phys. Chem. C 123, 18303 (2019).
\bibitem{Santos-2022} C.-K. Hu, J. Qiu, P. JP. Souza, J. Yuan, Y. Zhou, L. Zhang, J. Chu, X. Pan, L. Hu, J. Li, Y. Xu, Y. Zhong, S. Liu, F. Yan, D. Tan, R. Bachelard, C. J. Villas-Boas, A. C. Santos, and D. Yu, Quantum Sci. Technol. 7, 045018, (2022).
\bibitem{R.Zheng-2022} R.-H. Zheng, W. Ning, Z.-B. Yang, Y. Xia, and S.-B. Zheng, New J. Phys. 24, 063031 (2022).
\bibitem{Y.Zheng-2023} Y. Y. Ge, X. M. Yu, W. Xin, Z. M. Wang, Y. Zhang, W. Zheng, S. X. Li, D. Lan, and Y. Yu, Appl. Phys. Lett. 123, 154002 (2023).
\bibitem{Gemme-2024} G. Gemme, M. Grossi, S. Vallecorsa, M. Sassetti, and D. Ferraro, Phys. Rev. Research 6, 023091 (2024).
\bibitem{Yang-2024} F.-M Yang and F.-Q. Dou Phys. Rev. A 109, 062432, (2024).
\bibitem{Dou-2020} F. Q. Dou, Y. J. Wang, and J. A. Sun, EPL (Europhysics Letters) 131, 43001 (2020).
\bibitem{Dou-2022} F. Q. Dou, Y. J. Wang, and J. A. Sun, Frontiers of Physics 17, 31503 (2022).
\bibitem{Albash-2012} T. Albash, S. Boixo, D. A. Lidar, and P. Zanardi, New J. of Phys. 14, 123016 (2012).
\bibitem{Kato-1950} T. Kato, J. Phys. Soc. Japan. 5, 435 (1950).
\bibitem{Lidar3-2009} D. A. Lidar, A. T. Rezakhani, and A. Hamma, Adiabatic approximation with exponential accuracy for manybody
    systems and quantum computation, J. Math. Phys. 50, 102106 (2009).
\bibitem{Amin1-2009} M. H. S. Amin, Consistency of the adiabatic theorem, Phys. Rev. Lett. 102, 220401 (2009).
\bibitem{Steffen-2003} M. Steffen, W. V. Dam, T. Hogg, G. Breyta, and I. Chuang, Experimental implementation of an adiabatic
    quantum optimization algorithm, Phys. Rev. Lett. 90, 067903 (2003).
\bibitem{Lidar1-2005} M. S. Sarandy and D. A. Lidar, Adiabatic quantum computation in open systems, Phys. Rev. Lett. 95, 250503 (2005).
\bibitem{Lidar2-2005} M. S. Sarandy and D. A. Lidar, Adiabatic approximation in open quantum systems, Phys. Rev. A 71, 012331 (2005).
\bibitem{Lidar-2015} T. Albash and D. A. Lidar, Decoherence in adiabatic quantum computation, Phys. Rev. A 91, 062320 (2015).
\bibitem{Born-1928} M. Born and V. Fock, Z. Phys. 51, 165 (1928).
\bibitem{Farhi-2001} E. Farhi, J. Goldstone, S. Gutmann, J. Lapan, A. Lundgren, and D. Preda, Science 292, 472 (2001).
\bibitem{Johnson-2011} M. W. Johnson et al,  Nature. 473, 194 (2011).
\bibitem{Child-2001} A. M. Childs, E. Farhi and J. Preskill, Phys. Rev. A 65, 012322 (2001).
\bibitem{Teufel-2003} S. Teufel, Adiabatic Perturbation Theory in Quantum Dynamics, (Berlin: Springer, 2003).
\bibitem{Sarandy-2005} M. S. Sarandy, and D. A. Lidar, Phys. Rev. A 71, 012331 (2005).
\bibitem{Alicki-2006} R. Alicki, D. A. Lidar and P. Zanardi, Phys. Rev. A 73, 052311 (2006).
\bibitem{Joye-2007} A. Joye, Commun. Math. Phys. 275, 139, (2007).
\bibitem{Huang-2008} X. L. Huang, X. X. Yi, C. Wu, X. L. Feng, S. X. Yu, and C. H. Oh, Phys. Rev. A 78, 062114 (2008).
\bibitem{Amin-2009} M. H. S. Amin, C. J. S. Truncik, and D. V. Averin, Phys. Rev. A 80, 022303 (2009).
\bibitem{Oreshkov-2010} O. Oreshkov, and J. Calsamiglia, Phys. Rev. Lett. 105, 050503 (2010).
\bibitem{Salmilehto-2010} J. Salmilehto J, P. Solinas, J. Ankerhold and M. Mttonen, Phys. Rev. A 82, 06211 (2010).
\bibitem{Marangos-1998} J. P. Marangos, J. Mod. Opt. 45, 471 (1998).
\bibitem{Bergmann-1998} K. Bergmann, H. Theuer, and B.W. Shore, Rev. Mod. Phys. 70, 1003 (1998).
\bibitem{Demirplak-2003} M. Demirplak and S. A. Rice, J. Phys. Chem. A 107, 9937-9945 (2003).
\bibitem{Fleischhauer-2005} M. Fleischhauer, A. Imamoglu, and J. P. Marangos, Rev. Mod. Phys. 77, 633 (2005).
\bibitem{Lindblad-1976} G. Lindblad, Commun. Math. Phys. 48, 119 (1976).
\bibitem{Martinis-2003} J. M. Martinis, S. Nam, J. Aumentado, K. M. Lang, and C. Urbina, Phys. Rev. B 67, 094510 (2003).
\bibitem{Paraoanu-2011} J. Li, G. S. Paraoanu, K. Cicak, F. Altomare, J. I. Park, R. W. Simmonds, M. A. Sillanpaar, and P. J. Hakonen, Phys. Rev. B 84, 104527 (2011).
\bibitem{Sillanpr-2012} J. Li, M. A. Sillanpaar, G. S. Paraoanu, and P. J. Hakonen, J. Phys.: Conf. Ser. 400, 042039 (2012).
\bibitem{Allahverdyan-2004} A. E. Allahverdyan, R. Balian and T. M. Nieuwenhuizen, Maximal work extraction from
finite quantum systems. Eur. Phys. Lett 67, 565 (2004).
\bibitem{Devoret-2013} M. H. Devoret and R. J. Schoelkopf, Science 339, 1169 (2013).
\bibitem{Wendin-2017} G. Wendin, Rep. Prog. Phys. 80, 106001 (2017).
\bibitem{Gu-2017} X. Gu, A. F. Kockum, A. Miranowicz, Y. X. Liu, and F. Nori, Phys. Rep. 718719, 1 (2017).
\bibitem{Koch-2007} J. Koch, T. M. Yu, J. Gambetta, A. A. Houck, D. I. Schuster, J. Majer, A. Blais, M. H. Devoret, S. M. Girvin, and R. J.
    Schoelkopf, Phys. Rev. A 76, 042319 (2007).
\bibitem{You-2007} J. Q. You, X. Hu, S. Ashhab, and F. Nori, Phys. Rev. B 75, 140515(R) (2007).
\bibitem{Barends-2013} R. Barends, J. Kelly, A. Megrant, D. Sank, E. Jeffrey, Y. Chen, Y. Yin, B. Chiaro, J. Mutus, C. Neill, P. OMalley, P. Roushan, J. Wenner, T. C. White, A. N. Cleland, and J. M. Martinis, Phys. Rev. Lett. 111, 080502 (2013).
\bibitem{Lidar-2009} D. A. Lidar, A. T. Rezakhani, and A. Hamma, Adiabatic approximation with exponential accuracy for manybody
    systems and quantum computation, J. Math. Phys. 50, 102106 (2009).
\bibitem{Bhatia-1997} R. Bhatia,  Graduate Texts in Mathematics vol 169 Matrix Analysis, (New York: Springer) (1997).
\bibitem{Lidar4-2008} D. A Lidar, P. Zanardi, and K. Khodjasteh, Phys. Rev. A 78 012308 (2008).
\end{thebibliography}
\end{document}